\documentstyle[12pt,a4,epsf]{article}

\makeatletter
\@addtoreset{equation}{section}
\makeatother

\newcommand{\VEV}[1]{\left\langle #1 \right\rangle}

\begin{document}
\begin{titlepage}

\begin{flushright}
KUNS-1660\\
hep-ph/0004260\\
\today
\end{flushright}

\vspace{4ex}

\begin{center}
{\large \bf
Vacuum selection by recollapsing
}

\vspace{6ex}

\renewcommand{\thefootnote}{\alph{footnote}}

Nobuhiro Maekawa
\footnote
{
e-mail: maekawa@gauge.scphys.kyoto-u.ac.jp
}

\vspace{4ex}
{\it Department of Physics, Kyoto University,\\
     Kyoto 606-8502, Japan}\\
\end{center}

\renewcommand{\thefootnote}{\arabic{footnote}}
\setcounter{footnote}{0}
\vspace{6ex}

\begin{abstract}
We discuss the possibility that the vacuum is dynamically determined
in the history of the universe. The point is that some of the bubbles with
a certain vacuum shrink by the evolution of the universe via gravity and 
may become black holes. 
When the temperature of the phase transition $T_{PT}$ is
higher than $10^9$ GeV, these black holes evaporate until now. If $T_{PT}<10^9$
GeV, we may see these black holes in our universe.
It is interesting that in many cases false vacua are favored in the context
of cosmology. By using this argument, supersymmetry(SUSY), if it exists, 
can be broken
cosmologically. We can guess the SUSY breaking scale from the mass
of the black holes.
\end{abstract}
\end{titlepage}

\section{Introduction}
It is well known that the cosmology is strongly related with the particle
physics.
\cite{Kolb} 
The cosmology gives various constraints to the particle physics. 
For examples, the present ratio of the radiation density and the critical
density gives constraints to the mass and the interaction of the stable 
particles and the 
success of the Big-Bang Nucleosynthesis requires shorter lifetime 
than about one second for the unstable particles with large number density.
On the other hand, elementary particle physics also gives various effects
to the cosmology. In order to examine the early universe, we need the 
knowledge of the elementary particle physics, because the evolution of the 
universe is strongly dependent on the physics at the small scale. More
concretely the elementary particle theory gives some candidates for the 
dark matter and the concept of the field theory is applied to solve some 
cosmological problems as in inflation.
\cite{Linde}

Are there any possibility that the present elementary particle theory
is more dynamically determined by the history of the universe?
This is the point which we examine in this paper.

Before discussing cosmological situation,
let us roughly recall the theory of evolution for creatures. 
Every creatures have genes. The genes essentially determine 
the behavior of the creatures. The circumstance around the creatures 
select the creatures by their behaviors. If some creatures behave 
more suitably for their circumstance, then the creatures have longer life 
and create more children and their genes.
Namely genes with longer life survive in the history.

What we would like to do is to apply the above logic to the 
elementary particle physics. The elementary particle theories
determine the evolution of the universes. Some of the universes are 
expanding and some of them are recollapsing. As the result, the 
universes with longer life survive. In other words, theories which
make the universe have longer life survive. What are the features
of theories with long or short life? This is main motivation in 
this paper.

But what does 'many' universes mean? The answer in this paper is
that it means many bubbles with different vacua which appear 
at the phase transition.
In the early universe, there must be several phase transitions. If there are
several minima of the potential at the phase transition, there may be several
kinds of bubbles with different vacua in the universe simultaneously. 
Since generally the bubbles with different field theories evolve differently,
the bubbles with different vacua may evolve differently. 
Some of them may expand more rapidly than the other bubbles and dominate the 
other universes, and some of them may shrink and become black holes when the 
radius of the bubbles becomes smaller than the Schwarzschild radius.
As the result, we have 
many universes with different lifetimes and the selection of the vacua happens.

The vacuum selection by using inflation has been argued in the literature.
\cite{Linde,eternity,Izawa}
A. Linde argued that the initial condition of the vacuum expectation value
of the scalar is naturally selected so that the inflation happens, because
inflated universe dominates the universe. After his argument, various features
have been discussed in the context of the inflation. However, in a sense
this selection is not true selection, because the universes without inflation
also survive unless the universes are recollapsing, though they occupy only 
smaller region in the whole universe. 
Therefore it is possible that we are living in one of these universes,
though the simple argument of probability teaches us that we need the strong
luckiness or the anthropic principle.

On the other hand, as long as we know, 
there are only a few papers on the vacuum selection by using recollapsing via
gravity force.
\cite{Banks}
Banks et al. have argued this possibility in the context of the string moduli 
problem. They argued that the universes
with negative cosmological constant shrink following Friedmann equation.
By this argument, they insisted that the universes with negative cosmological
constant are unstable.

We think that it is important to examine more generally the features of 
universes which shrink by gravity effect. By 
studying on the universes with short lifetime, we can understand the features 
of universes with long
life and the features of our particle field theories which we have in our
universe. 
In this paper, under simple assumptions we examine several cases in which
the vacuum selection by recollapsing happens. It is interesting that in many
cases false vacua are selected by this argument. Namely, cosmology prefers
local minima. And we insist that the recollapsing bubbles become black holes.
If the black holes evaporates until now, this selection by recollapsing becomes
true selection. In other words, the bubble with the special vacuum is too weak 
to survive. It is also interesting that we may be able to observe the other 
universes as the 
black holes in our horizon unless the black holes evaporate until now. 
Moreover, we discuss the application of the cosmological vacuum selection
to the elementary particle theories, especially to the cosmological 
supersymmetry(SUSY) breaking.

The plan of this paper is as follows.
After the review on the evolution of the universe in section 2, we examine 
several cases in which the selection by recollapsing happens in section 3.
In section 4, we discuss various features on the primordial black holes 
which are formed by the recollapsing via gravity force. 
In section 5, we apply the result that cosmology prefers 
the local minima to the particle theory. Especially we examine the possibility
that the SUSY is cosmologically broken. Finally we discuss 
our assumptions and summarize our arguments.

\section{Evolution of the Universe}
If the universe is homogeneous and isotropic, the metric can be written in the
form
\begin{equation}
ds^2=dt^2-a(t)^2\left( \frac{dr^2}{1-kr^2}+r^2d\theta^2+r^2\sin^2\theta
d\phi^2\right),
\label{RW}
\end{equation}
which is Robertson-Walker metric. Under this metric, the Einstein 
equation
\begin{equation}
G_{\mu\nu}\equiv R_{\mu\nu}-\frac{1}{2}R g_{\mu\nu}
=8\pi G T_{\mu\nu}+\Lambda g_{\mu\nu},
\label{Einstein}
\end{equation}
where $G_{\mu\nu}$, $R_{\mu\nu}$, $T_{\mu\nu}$ and $\Lambda$ are the Einstein
 tensor,  the Ricci tensor, the stress-energy tensor for all the fields and 
 a cosmological constant, respectively, leads to the Friedmann equation
\begin{equation}
H^2\equiv \frac{\dot a^2}{a^2}=-\frac{k}{a^2}+\frac{8\pi G}{3} 
\rho \label{Friedmann} 
\end{equation}
and the energy equation
\begin{equation}
d(\rho a^3)=-p d(a^3),
\label{energy}
\end{equation}
where $H$ is the Hubble constant.
Here we take
\begin{eqnarray}
\rho&=&\bar \rho + \rho_\Lambda=\rho_R+\rho_M+\rho_\Lambda \\
p&=&\bar p+p_\Lambda=p_R+p_M+p_\Lambda \\
T^\mu_\nu&=&{\rm diag}(\bar \rho, -\bar p, -\bar p,-\bar p) \\
\rho_\Lambda&=&-p_\Lambda\equiv\frac{\Lambda}{8\pi G},
\end{eqnarray}
where $\rho_{(R,M)}$ and $p_{(R,M)}$ are the energy density and pressure for
(relativistic fields(radiation), non-relativistic fields(matter)), respectively.
Though the energy density $\rho$ and the pressure $p$ are the sum of the 
contribution from the relativistic fields, non-relativistic fields and
the cosmological constant, the energy equation is realized independently
if the energy transfer among relativistic fields, non-relativistic fields
and the cosmological constant is negligible. Namely
\begin{equation}
d(\rho_A a^3)=-p_A d(a^3),
\label{rho-p}
\end{equation}
where $A=R,M,\Lambda$. When
\begin{equation}
p_A=\gamma_A \rho_A,
\end{equation}
the differential equation (\ref{rho-p}) can be solved as
\begin{equation}
\rho \propto a^{-3(1+\gamma_A)}.
\end{equation}
Here $\gamma_R=1/3$, $\gamma_M=0$ and $\gamma_\Lambda=-1$.
Then the Friedmann equation is rewritten as
\begin{eqnarray}
-\frac{1}{2}k&=&\frac{1}{2}\dot a^2+V(a) \\
V(a)&=&-\frac{4\pi G}{3}a^2\rho (a),
\end{eqnarray}
which has the same kinematics as of one dimensional particle moving with 
the energy
$E=-k/2$ under the potential $V(a)$. In the following analysis, we take the
initial condition $\dot a>0$, namely the universe is expanding in the 
beginning.

When $\rho_\Lambda=0$, the global behavior is determined by the signature of 
the parameter $k$. If $k$ is positive, the universe becomes recollapsing. If
$k$ is zero or negative, the universe keeps expanding. 

On the other hand, if the vacuum energy dominates the energy density, the 
signature of the cosmological constant plays essential roles. 
If the cosmological constant is positive, the universe is expanding 
exponentially. This situation is called inflation. If the cosmological constant
is negative, the universe becomes recollapsing.

Let us estimate the time scale for recollapsing by solving the Friedmann 
equation (2.12). With vanishing cosmological constant, the time scale 
$\tau_{(R,M)}$
 for the (Radiation, Matter) dominated case is calculated
as
\begin{eqnarray}
\tau_R &=& \frac{1}{H_I}\left( \frac{2}{\Omega_I-1}+O(1)\right) = 
         \frac{O(t_I)}{(\Omega_I-1)} \\
\tau_M &=& \frac{1}{H_I}\left( \frac{\pi}{(\Omega_I-1)^{3/2}}
+O((\Omega_I-1)^{-1})\right)=\frac{O(t_I)}{(\Omega_I-1)^{3/2}}
\end{eqnarray}
where $\Omega\equiv \rho/\rho_c$ is defined by using critical density 
$\rho_c\equiv 3H^2/(8\pi G)$ and the suffix $I$ represents the initial value
at the time $t_I$. We can easily find that
when $\Omega-1=O(1)$, the time scale for recollapsing is of order 
$t_I\sim 1/H_I$. This estimation show that the universe recollapses soon
after $\Omega-1$ becomes of order 1.

With the negative cosmological constant, the time scale $\tau_{ADS}$ for 
recollapsing is
\begin{equation}
\tau_{ADS}\sim 2\int_{a_I}^{a_{max}} da \frac{1}{\sqrt{-k-2V(a)}}
            =\frac{1}{H_I\sqrt{-\Omega_I}}\left( \pi-2 
           {\rm Arcsin}\left(\sqrt{\frac{-\Omega_I}{1-\Omega_I}}\right)\right)
           =O(t_I),
\end{equation}
where $a_{max}$ is determined by the relation 
$-k=2V(a_{max})$.
This means that the universe recollapses soon after the vacuum energy 
dominates the energy density of the universe.
The results in this section are summarized in table 1.

\vspace{5mm}
\begin{center}
\begin{tabular}{|c|c|c|c|} 
\hline
 & $k>0(\Omega>1)$ & $k=0(\Omega=1)$ & $k<0(\Omega<1)$ \\ \hline
\hline
$\Lambda>0$ & - & inflation & inflation \\ \hline
$\Lambda=0$ ($\rho=\rho_R$) & $O(t_I)(\Omega-1)^{-3/2}$ & $\infty$
 & $\infty$
\\ \hline
$\Lambda=0$ ($\rho=\rho_M$) & $O(t_I)(\Omega-1)^{-1}$ & $\infty$ & $\infty$ 
\\ \hline
$\Lambda<0$ & $O(t_I)$ &$(O(t_I))$ & $(O(t_I))$ \\ \hline
\end{tabular}

\vspace{0.5cm}
{\rm table 1. The lifetime for recollapsing}
\end{center}

\section{Examples}
In this section we discuss several examples in which collapsing bubbles
and expanding bubbles exist simultaneously, namely vacuum selection by
recollapsing occurs.

It is not so easy to understand the evolution of the universe at the phase
transition, because generally the universe is not homogeneous nor isotropic.
It is also difficult to solve the Einstein equation in such general cases,
therefore we take several assumptions.

The first assumption is that the Friedmann equation can be used for estimating
the evolution of the bubble universes. If the size of the bubbles
is larger than the horizon length $d_H\sim 1/H$ and the space inside the 
bubble is almost homogeneous and isotropic,
then it is natural to use the Friedmann equation for the evolution of the 
space inside the bubble.
Such calculation is adopted by several people who discussed the evolution of 
the universe,
\cite{Linde,eternity,Izawa,Banks}
for example, in the context of the chaotic inflation by A. Linde.
Here we use the Friedmann equation for the evolution of the bubble 
universes.

The second assumption is that the phase transition does not change the energy
density. The phase transition can change only the carrier of the energy among
relativistic particles, non-relativistic particles and vacuum. 
This is only for the simplicity.

We consider the following situation.
At the higher
temperature, there is only one minimum, which disappears and two new minima
appear around the critical temperature $T_{PT}$. Suppose that the two phases
coexist for a while after the phase transition and the time scale of the
coexistence is larger than the typical time scale for recollapsing, e.g.,
$O(t_I)$.
Under these assumptions, we examine several cases in which collapsing bubbles
and expanding bubbles exist at the same time. 

The first case is that the universe has the larger energy density than the
critical density($k>0$), therefore $\Omega_{PT}>1$ at the time $t_{PT}$ 
when the phase transition occurs.
After the phase transition, the potential has one global minimum with 
vanishing cosmological 
constant and the other minimum which is local. The bubble universe
with the true vacuum recollapses because the energy density is 
larger than
the critical density. The time scale for the recollapsing is 
$O(t_I)$ unless the energy density is tuned to the critical density.
It is natural that the recollapsing bubbles become
black holes, which we will discuss later. On the other hand, the bubble 
universe with the false 
vacuum inflates if the cosmological constant dominates the energy
density before reaching the time when the bubble begins to shrink.

In order to regard the inflating universe as our universe, the inflation
must stop some time because our universe does not inflate at present, namely 
cosmological constant must vanish till now. Here we expect that there is
an unknown mechanism which realizes the vanishing cosmological constant.
And such a mechanism automatically stop the inflation and the vacuum energy
is released in the universe(reheating). Of course we have no reliable 
mechanism for the vanishing cosmological constant, so we do not discuss
more on how to stop the inflation here.

The second case is that the energy density is smaller than the critical
density. After the phase transition, the local minimum has
vanishing cosmological constant, and the global minimum has negative
cosmological constant.
After the negative cosmological constant in the
bubble universe of the true vacuum dominates the energy density of the universe,
 the bubble 
universes begin to shrink. Therefore the bubble universes with true vacuum 
become black holes and the bubble universe with the false vacuum keeps
expanding and 
dominates our universe.

In the above two cases, our universe is on the false vacuum.
The false vacuum decays to the true vacuum by quantum tunneling effect.
\cite{Linde2,Coleman}
The lifetime
of the universe must be longer than the age of the present universe if you 
would like to regard the universe as our universe. 
We will shortly discuss the 
lifetime of the universe.

Of course, we can think about the intermediate situations between the above two
cases. Namely, the potential has a global minimum with the negative cosmological
constant and a local minimum with the positive cosmological constant. In this 
case, the bubble universes with the true vacuum become black holes and
the bubble universes with the false vacuum inflate. 
Notice that the above vacuum selection happens even if the initial number 
density of the bubble universes with the false vacuum is much smaller than
that of the true vacuum. Since the bubble universes with true vacuum are 
recollapsing, the tiny part of the universe with the false vacuum can dominate 
the whole universe
if it exists.

The final case is that the energy density is almost the same as the critical 
density, but a little bit larger than the critical density. And the both vacua
are degenerate with the vanishing cosmological constant. Therefore, bubbles with
both vacua turn to recollapse some time. Moreover, we take the situation that
one of the vacua gives the matter dominated universe, and the other gives the 
radiation dominated universe. 
Here for simplicity, soon after the phase transition, such a difference is 
realized.
Then the time scales for recollapsing are different and dependent on whether
the bubble are the radiation dominant or matter dominant. As we discussed the
 lifetime in the 
previous section, the matter dominated universe has longer lifetime than the
radiation dominated universe under the same initial conditions. 
If the energy density is near the critical density,
the difference becomes large, because the lifetime is proportional to 
$1/(\Omega-1)$ for the radiation dominated universe and $1/(\Omega-1)^{\frac{3}
{2}}$ for the matter dominated universe. If the radiation dominated universe 
become black hole with shorter period than the age of our universe and the
matter dominated universe has longer lifetime than the age of our universe, then
the vacuum selection is realized. Since $\Omega \sim 1$ is realized after
inflation, it may be possible that even in our real world, the situation 
happens.

By examining the above cases and the other cases of vacuum selection
\cite{Linde,Izawa} by 
inflation, we can easily conclude that the cosmology prefers local minima.
However, in the ordinary field theory, the local minimum is unstable, namely
the local minimum decays to the global minimum. How long is the lifetime
of the unstable vacuum? 
In the end of this section, we discuss the lifetime of the 
local minimum by the instanton calculation,
\cite{Linde2,Coleman,DeLuccia}
which gives
\begin{equation}
\tau \sim \frac{1}{V\Lambda^4}e^{S(\bar \phi)},
\end{equation}
where $V$ and $\Lambda$ are the volume of the universe and the typical scale
of the potential of the scalar $\phi$. Here $\bar \phi$ is an instanton 
solution and $S(\phi)$ is the action of the scalar field $\phi$.
Since the lifetime of the local minimum should be longer than the age
of our universe $\tau_0\sim 10^{20}$ seconds, we get the condition
\begin{equation}
10^{200}\frac{\Lambda^4}{(GeV)^4} \leq e^{S(\bar \phi)}.
\end{equation}
When the difference of the potential energy between the two minimum is
much smaller than the typical scale $\Lambda^4$, then we get 
\begin{equation}
S(\bar \phi)\sim 2 \pi^2 O(1)/\epsilon^3,
\end{equation}
where $\Delta V \sim \epsilon \Lambda^4$.
Therefore $\epsilon \sim 1/10$ is small enough to satisfy the above condition.
This situation seems not to require strong fine tuning and is not so 
difficult to be satisfied. It is not natural that a scalar potential
has no local minimum with the above feature. Therefore we think that 
it is not unnatural 
that we are living in a false vacuum.

\section{Primordial black hole}
The black hole which is induced in the early universe is called primordial 
black hole. In our situation, some bubble universes may become the primordial
black holes. But how heavy primordial black holes are induced and how many
black holes are there in our universe? 
Generally it is not so easy to give a concrete conclusion on the formation
of the black holes at the phase transition. In the literature, the 
formation of the black holes has been discussed in the various cases with 
various assumptions.
\cite{Sato,collide,Hall,numerical} 
However, for the situation that the size of coexisting bubbles is around 
the horizon
length $d_H \sim O(1/H)$, which can be reached by the light from the beginning
of the universe,
we can make a simple argument.
It is consistent with the previous assumption that Friedmann equation can be 
applied to the evolution of the bubble universes.

The mass of the black holes can be estimated from the size of the bubbles
and the energy density. 
 The energy density is given by
\begin{equation}
\rho\sim T_{RC}^4,
\end{equation}
where $T_{RC}$ is the temperature at which the bubble begin to recollapse.
Since in the most cases we discussed $T_{RC}$ is the same order of the 
temperature of the phase 
transition $T_{PT}$, we use $T_{PT}$ in the following argument.
We can estimate the order of the mass of the primordial black
hole by using the temperature and Planck scale $M_P$ as
\begin{equation}
M_{BH}\sim d_H^3 \rho\sim M_P^3/T_{PT}^2.
\end{equation}
Here we use the relation $H^2 \sim G\rho$.
The Schwarzschild radius $R_S=2GM_{BH}$ for the black hole becomes
\begin{equation}
R_S\sim \frac{1}{H}.
\end{equation}
It is well-known but surprising that the Schwarzschild radius is the 
same order of the 
horizon length which is the size of the bubbles.
Therefore it is expected that the shrinking bubbles become 
black holes soon
after the density outside the bubble becomes smaller than the inside
one. It is natural that the recollapsing bubbles become black holes.

On the other hand, black holes are evaporating by Hawking radiation.
\cite{Hawking}
The lifetime is given by
\begin{equation}
\tau_{BH}\sim \frac{2560\pi}{g_*}G^2M_{BH}^3\sim \frac{2560\pi}{g_*}
   \frac{M_P^5}{T_{PT}^6}.
\end{equation}
Here $g_*$ is the degrees of freedom of the particles at the Hawking 
temperature $T_H=1/(8\pi GM_{BH})$.
From the table 2

\vspace{5mm}

\begin{center}
\begin{tabular}{|c|c|c|c|c|c|} 
\hline
$T_{PT}$ ({\rm GeV}) & 1 & 100 & $10^9$ & $10^{12}$ & $10^{19}$ 
 \\ \hline
$M_{BH}$ (g)& $O(10^{33})\sim O(M_{Solar})$ & $O(10^{29})$ &
$O(10^{15})$  & $O(10^9)$ & $O(10^{-5})$  \\ \hline
$T_H({\rm GeV})$ & $O(10^{-21})$ & $O(10^{-17})$ & $O(10^{-3})$
 & $O(10^3)$ & $O(10^{17})$ \\ \hline
$\tau_{BH}$ (s) & $O(10^{74})$ & $O(10^{62})$ & $O(10^{20})\sim \tau_0$
 & $O(1)$ & $O(10^{-42})$ 
 \\ \hline
\end{tabular}

\vspace{5mm}
table 2. Here $M_{Solar}$ is the solar mass and $\tau_0$ is the age of our
universe.
\vspace{5mm}
\end{center}

\noindent
we can easily find that the primordial black holes may be seen in our universe
if the temperature of the phase transition is less than $10^9$ GeV. This is
because the age of our universe is around $10^{20}s$. The black holes caused
by the phase transition above $10^9$ GeV evaporate until now,
though there may exist some remnant of the black holes. 
Notice that
the mass of the black hole from the QCD phase transition is around the solar
mass $M_{Solar}$ which is the same order as the mass of the massive compact 
halos objects
(MACHO) found by using gravity
lensing effect.
\cite{MACHO}

The number density is difficult to be estimated, because it is strongly 
dependent on the expanding rate of our universe, the shape of the potential,
the features of the phase transition and the initial conditions. But here
we roughly estimate the number density, though this estimation is not so
reliable in general cases. 
The number density at the phase transition can be naively estimated as
\begin{equation}
n_{BH}(t_{PT})\sim \frac{1}{d_H^3} \sim H^3 \sim \frac{T_{PT}^6}{M_P^3}.
\end{equation}
Since the black holes are regarded as the non-relativistic objects, 
the number density after
the phase transition is
\begin{equation}
n_{BH}(t) = n_{BH}(t_{PT})\left( \frac{R(t_{PT})}{R(t)}\right)^3.
\end{equation}
In the case 2, the density of the black holes dominates the energy density
soon, which means that the universe becomes matter dominant. 
Since usually we think that the temperature of the phase transition is much 
larger than $T\sim 5.5$ eV at which matter density becomes almost equal to 
the radiation density,
in order to regard the universe as our universe, we need
the evaporation of the black holes(i.e., the scale of the phase transition
is larger than $10^9$ GeV) or another inflation realizing smaller density
of the black holes. 

In the case 1, after the inflation ends, the number density is estimated as
\begin{equation}
n_{BH}(t)=n_{BH}(t_{PT})e^{-3Ht_E}\left( \frac{R(t_E)}{R(t)}\right)^3,
\end{equation}
where $t_E$ is the time when the inflation ends. 
Unfortunately in this case 
it is almost impossible to see the black holes in 
our universe. This is because there is only a few black hole at the most in 
our universe within the present horizon since the present universe is almost
flat.

The interesting case is the intermediate case between cases 1 and 2. Namely
the local minimum has a positive cosmological constant and the global minimum
has a negative cosmological constant. Suppose that before the phase transition,
$\Omega=1$ has already been realized
by another inflation etc. In this case, the bubbles with the global minimum 
shrink and become black holes soon, and the bubbles with the local minimum is 
inflating. The number density can be estimated as in the eq. (4.7), 
and we can realize
any number density by tuning the time the inflation stops. Therefore we may see
the primordial black holes caused by the phase transition and recollapsing 
process in our universe. In this scenario, the primordial black holes can be
dark matter candidates for any level. Though it is interesting to examine the 
possibility that the small primordial black hole becomes the dark matter, 
it is beyond the subject in this paper.

\section{Cosmological SUSY breaking}
In this section, we examine applications of the above results to the particle 
physics. If it is usual that the local minimum is selected dynamically by 
disappearance of the global minimum, it must be interesting to examine the 
possibility that we are living in the local minimum more 
seriously. Of course, 
such a possibility has been already examined, because the local minimum can 
have longer lifetime than the age of our universe. But for the most of people,
such a passive reason seems not to be enough to take the possibility seriously.
Actually, many people have discussed the conditions for realizing the 
expected vacuum as the global minimum
\cite{Casas}
 and only the structure of the space of the
global minimum (moduli space) in the 
supersymmetric field theories. However, as we argued, the universe with the 
global minimum may collapse and disappear in the history of the universe.
Though the results is dependent on the initial conditions, it must be 
important to examine the possibility that we are living in the local minimum
because we may have to stay on the local minimum.
The examination of the possibility may be crucial for finding
the real vacuum. For example, in the string theory, to realize the dilaton 
stability is one of the interesting problem to be solved. By changing the
K\"ahler potential of the dilaton, we can stabilize the dilaton vacuum
though
this vacuum is often at the local minimum. If the local minimum is 
cosmologically selected, we do not have to mind why we are not living in the 
global minimum, which is too weak to survive. Moreover we think it important to 
examine the 
possibility in various phase transition, for examples, in the QCD phase 
transition,
the electroweak phase transition and GUT phase transition. But here we discuss
only the SUSY breaking.

If the nature has SUSY, scalars must exist.
It seems to be natural that the scalar potential has several local minima which
have features for longer life than the age of our universe. Since cosmology
prefers local minimum, it is natural that such local minima is selected in the
history of the universe. Since only the global minimum keeps the SUSY in the 
context of the global SUSY theories, in the theory with local minimum SUSY is 
spontaneously broken. Notice that SUSY is spontaneously broken even if 
SUSY vacua exist. Therefore SUSY is broken by cosmology. 

Where is the natural scale for the SUSY breaking? If a theory has SUSY at 
a scale, then the theory has generally a scalar potential. It is natural that
the potential has the local minima which break SUSY. The natural SUSY breaking 
scale is the scale. If the theory
has a SUSY at the Planck scale, then it is natural that SUSY is broken at the 
Planck scale.
Of course, such an argument does not reject the low energy SUSY. 
The theory may have no local minima with long life. Even if it has local minima,
the initial conditions may allow the universe with vanishing cosmological 
constant to expand. 

By using cosmological SUSY breaking, we can make simpler models in which SUSY
is spontaneously broken. The point is that these models are allowed to have
SUSY vacua. Even if the models have SUSY vacua, cosmology prefers SUSY breaking
local minima. We examine a simple model with gauge mediated SUSY breaking.

Several years ago, Izawa-Yanagida
\cite{IY} 
and Intriligator-Thomas
\cite{Thomas}
 proposed one of the simplest dynamical
SUSY breaking model which has a SUSY SU(2) gauge group with four doublet 
chiral superfields $Q_i (i=1,\cdots, 4)$ and several singlets $S$ and $S^a 
(a=1,\cdots,5)$. Here the suffix $i$ is a flavor index and $a$ is the index
of five dimensional representation of SP(4) global symmetry which they adopted
as the global symmetry of the superpotential
\begin{equation}
W=ySQQ+\bar yS^a(QQ)_a,
\end{equation}
where $(QQ)$ and $(QQ)_a$ are one and five dimensional representations of 
SP(4) which are formed from a suitable combination of $Q_iQ_j$.
The effective superpotential is given by
\begin{equation}
W_{eff}=y \Lambda SV+\bar y \Lambda S^aV_a
\end{equation}
with constraint $V^2+V_a^2-\Lambda^2=0$. Here the composite meson fields
$V\sim (QQ)/\Lambda$ and $V_a\sim (QQ)_a/\Lambda$ are the low energy degrees 
of freedom and $\Lambda$ is a dynamical scale of the SU(2) gauge theory.
When $y<\bar y$, the condensation
\begin{equation}
\langle V \rangle = \Lambda, \langle V_a \rangle =0
\end{equation}
is realized at the global minimum, and
the effective potential is approximately rewritten as
\begin{equation}
W=y \Lambda^2 S.
\end{equation}
The F-term of the $S$ field $F_S$ is $y \Lambda^2\neq 0$, which means that
the SUSY is spontaneously broken. 
Notice that if the K\"ahler potential of the $S$ field is minimal one, 
then the vacuum
expectation value(VEV) of the $S$ field is undetermined, namely, the potential
has a flat direction. 

In order to mediate the SUSY breaking effect to the ordinary gauginos, squarks
 and 
sleptons via the standard model gauge interaction(gauge mediation),
\cite{gauge}
we 
introduce vector-like messenger fields $q$ and $\bar q$ which
has quantum numbers of the standard gauge group and the superpotential
\begin{equation}
W=\lambda S\bar q q.
\end{equation}
If the vacuum expectation values of these fields remain unchanged by
this deformation, namely,
\begin{equation}
\langle V \rangle = \Lambda, \langle V_a \rangle =0, 
\langle q \rangle = \langle \bar q \rangle =0,
\end{equation}
the supersymmetry is broken again.
If we take $\langle S \rangle^2 > F_S$,
then ordinary gauge mediation calculation gives 
the gaugino masses
\begin{equation}
M_a\sim\frac{\alpha_a}{4\pi}\frac{\VEV{F_S}}{\VEV{S}}
\end{equation}
via one loop diagram and scalar fermion masses 
\begin{equation}
\tilde m_f^2\sim\sum_a\left(\frac{\alpha_a}{4\pi}\right)^2
 \left(\frac{\VEV{F_S}}{\VEV{S}}\right)^2
\end{equation}
via two loop diagram. Here $\alpha_a\equiv g_a^2/(4\pi)$ and $g_a$ are the
gauge couplings of the standard model. Notice that the order of the SUSY 
breaking mass 
scales of sfermions is the same as of gauginos. This mediation mechanism 
can give 
the realistic mass pattern to the superpartners. 

This naive model is very simple but unfortunately by adding the above 
superpotential, SUSY vacua appear, namely the F-term equation
\begin{equation}
y\Lambda^2-\lambda \bar q q=0
\end{equation}
can be satisfied by taking non vanishing vacuum expectation value of
$\bar q q$. Therefore usually this model has not been examined as a 
dynamical SUSY breaking model. Several attempts to make the SUSY vacua
disappear have been made in the literature.
\cite{INY,Murayama}
However, since cosmology
prefers local minima, we do not mind that the SUSY vacua exist. What is 
important here is that the expected vacuum should be at least local minimum and
the lifetime of the vacuum is longer than the age of our universe.

The potential under the superpotential and the minimum K\"ahler potential
is
\begin{equation}
V=|y\Lambda^2-\lambda \bar q q|^2+|\lambda S|^2(|q|^2+|\bar q|^2)
 +{\rm D term}
\end{equation}
which has a flat direction along the direction $q=\bar q=0$.
Along the flat direction, when $|\lambda S|^2 > y \Lambda^2$, the mass
square matrix of the fields $q$ and $\bar q$ has only positive eigenvalues,
on the other hand, when $|\lambda S|^2 < y \Lambda^2$, it has negative
eigenvalues. Therefore the expected vacuum (5.6) is not even a local minimum.
However, generally, the K\"ahler potential can be deformed by quantum 
corrections
from the minimal one. Then the expected vacuum may be a local minimum.
Actually 
if the K\"ahler potential of the $S$ field is corrected as
\begin{equation}
K=|S|^2+\eta \frac{|y S|^4}{\Lambda^2} (\eta>0)
\end{equation}
by the strong gauge interaction,
then the effective potential along the flat direction $q=\bar q=0$ 
is given by
\begin{equation}
V_{eff} \sim |y|^2 \Lambda^4(1-4\frac{\eta}{\Lambda^2}|y|^4|S|^2).
\end{equation}
Since the effective potential is lifted in the large $|S|$ region because
the Yukawa coupling $y$ grows with the scale $S$,
\cite{Thomas,inverse}
the potential along the flat direction has a minimum. If the condition
$|\lambda S|^2 > y \Lambda^2$ is satisfied at the minimum, the minimum is
a local minimum in the whole potential.
Since cosmology prefers local minima, the local vacuum can 
be selected cosmologically if the lifetime is longer than the age of our
universe, which is strongly dependent on the parameters and the vacuum
expectation value.

This is a simple model in which cosmological SUSY breaking happens,
though we may need to introduce another singlet or non-renormalizable terms
in order to get the 
realistic scale of the supersymmetric Higgs mass term.

In this cosmological SUSY breaking scenario, bubbles with SUSY vacua shrink
and become black holes. It is interesting that the mass of the black holes
is determined by the SUSY breaking scale (see table 2). Therefore by observing
the small black holes in our universe, we may measure the SUSY breaking scale.
Since the SUSY breaking scale for the gauge mediation scenario is usually 
between $10^5$ GeV and $10^{10}$ GeV, the mass scale of the black holes is
$10^{13}{\rm g}< M_{BH} < 10^{23}{\rm g}$. 
Though the black holes with lighter mass than $10^{15}$ g evaporate till now,
it is interesting that the most of the gauge mediated SUSY breaking models 
($10^5 {\rm GeV}< \sqrt{F_S} < 10^9$GeV)
produce the small
black holes with longer lifetime than the age of our universe and
we can know the SUSY breaking scale from the mass of the black holes.

\section{Discussion}
Coleman and DeLuccia have argued on the probability of the first order phase 
transition including the gravity by finding the O(4) symmetric instanton 
solution of the scalar field theory with gravity. 
Since the solution has O(4) symmetry, the bubble wall velocity becomes
light velocity soon and the bubbles keep expanding. 
Even when the global minimum 
has negative cosmological constant, they found the O(4) symmetric solution,
therefore the bubble keeps expanding.
They concluded that the observer outside the bubble sees the 
bubble expanding, on the other hand, the space inside the bubble is collapsing.
Though this conclusion is strongly dependent on their assumption that
the O(4) solution gives the smallest action, this seems to give a counter 
example
of our assumption that the evolution of the bubble is determined by the 
Friedmann equation. Similar phenomenon is examined in a different situation 
at the first order phase
transition.
\cite{Sato} 
Of course including finite temperature effects, the wall
velocity may become much less than the light velocity(i.e., the solution
has not O(4) symmetry).
\cite{Linde3}
However, it suggests that our calculation may be too naive. More reliable 
calculation including numerical estimation is needed, which is a future subject.

The determination of the vacuum is equivalent to the determination of the 
theory in a sense, in some cases the determination of the couplings. If we can 
construct a scenario in which only very limited vacua survive in the evolution
of the universe, then we may be able to solve the fine tuning problems of 
couplings, scales et al., for examples, strong CP phase or Higgs mass, 
by using cosmology.
This direction may be interesting, but this is also one of the future subjects. 

The discussion in this paper is similar to the anthropic
 principle, which 
insists that physical parameters must be as they are, otherwise there does 
not exist creatures like human beings who think why physical parameters 
are like these.
There are some difference between the usual human principle and our arguments.
The biggest difference is that in our scenario other universes may be observed
in our universe as black holes. In this sense, we think that our argument is
similar to the theory of evolution for creatures, because we can observe a lot
of fossils (black holes) as the remnants of the extinct creatures (
recollapsing universes).

\section{Summary}

Under several assumptions we examined the several examples in which the 
vacuum is selected by 
recollapsing. It is interesting that the recollapsing universes become
black holes which may be observed in our universe. The mass of the
black holes 
can be roughly
estimated in the situation that the bubble radius is around the horizon
length, it is interesting that QCD phase transition may produce the 
black holes with solar mass, because MACHOs with solar mass have already 
been observed by gravity lensing effect. 

It is also interesting that the local minimum is preferable to global minimum.
If this result is applied to the SUSY theory, SUSY may be cosmologically 
broken
even if the theory has SUSY global vacua. We made a simple model in which 
SUSY is cosmologically broken. 

The assumptions we imposed in this paper may be too naive. But we believe
that even if the assumptions are not satisfied in many cases, the vacuum
selection by recollapsing can generally happen and tells us important clues
to know why our world is as it is.

In the sense of the theory of the evolution for creatures, it may be 
considerable that bubbles with different physical theories fight
each other and bubbles with theories which will make other bubbles 
disappear may survive. Such a consideration like the struggle for existence
may give clearer
features of the theories which survive the history of the universe.

\section{Acknowledgement}
The author expresses thanks to D. Wright for encouraging and collaboration 
at the
early stage of this work. He also thanks N. Sugiyama for valuable conversation
on MACHO experiments.
Part of this work has been done while he was visiting Technische Universit\"at
M\"unchen. He would like to thank the staffs, especially Prof. M. Lindner, 
the colleagues and
faculty of this institute for their hospitality. 

\end{document}